\documentclass{aa}  

\usepackage{graphicx}
\usepackage{txfonts}
\usepackage{siunitx}
\usepackage{lipsum}
\usepackage[normalem]{ulem}
\usepackage{makecell}
                                
\usepackage{lscape}             
                                
\usepackage{placeins}           
                                
\usepackage[table, dvipsnames]{xcolor} % for colors

\newcommand{\tripod}{\texttt{TriPoD}\xspace}
\newcommand{\tripodpy}{\texttt{TriPoDPy}\xspace}
\newcommand{\cudisc}{\texttt{cuDisc}\xspace}

\newcommand{\amax}{$a_{\rm max}$}
\newcommand{\OmegaK}{\Omega_\mathrm{K}}
\newcommand{\Msun}{\mathrm{M}_\odot}
\newcommand{\Lsun}{\mathrm{L}_\odot}
\newcommand{\Rsun}{\mathrm{R}_\odot}

\begin{document}

   \title{The influence of dust composition on accretion outbursts}

   \author{Nicolas~Kaufmann \inst{\ref{LMU}}, Anna~B.T.~Penzlin \inst{\ref{LMU}}, Alfie~Robinson \inst{\ref{imperial}}, Alexandros~Ziampras\inst{\ref{LMU},\ref{MPIA}}, Tilman~Birnstiel\inst{\ref{LMU},\ref{ORIGINS}},  Richard~Booth\inst{\ref{Leeds}}}
   
   \authorrunning{Kaufmann et al.}

   \institute{Ludwig-Maximilians-Universit{\"a}t M{\"u}nchen, Universit{\"a}ts-Sternwarte, Scheinerstr.~1, 81679 M{\"u}nchen, Germany\label{LMU}
   \and Astrophysics Group, Imperial College London, Prince Consort Road, London SW7 2AZ, UK\label{imperial}
    \and Max Planck Institute for Astronomy, K{\"o}nigstuhl 17, 69117 Heidelberg, Germany\label{MPIA}
    \and Exzellenzcluster ORIGINS, Boltzmannstr.~2, 85748 Garching, Germany\label{ORIGINS}
    \and School of Physics and Astronomy, University of Leeds, Leeds LS2 9JT, UK\label{Leeds}
   }

   \date{May 2026}

  \abstract

   {Episodic accretion outbursts have been shown to occur in protoplanetary discs, where they are caused by thermal instability at the inner edge of the dead zone. These outbursts periodically heat up the dead zone, which significantly alters the composition, chemistry and accretion onto the star.}
   {We investigate how these accretion outbursts affect the inner disc composition and how different dust compositions and properties affect the outburst dynamics.} 
   {We ran vertically integrated axis-symmetric dust and gas evolution models using the TriPoD method, including dust opacity-dependent heating and cooling, a dead-zone model, and compositional tracking of dust, including the evaporation and condensation of volatiles.}
   {When dust is considered to consist of a condensation sequence of silicates and volatiles, the outburst evaporates most of the dust within 0.5 au. The fast re-condensation and subsequent viscous evolution reset the disc between bursts. Additionally, we found that the burst cycle period and maximum accretion rate directly correlate with the dust sublimation temperature.}
   {}

   \keywords{protoplanetary disks – accretion, accretion disks — radiation: dynamics — methods: numerical — protoplanetary disks: composition}

   \maketitle
   \nolinenumbers

\section{Introduction}
The luminosity of young stellar objects shows a wide range of variable behaviour in magnitude and frequency \citep[for a review see][]{fischer_accretion_2023} that is thought to be associated with variable accretion onto the protostar \citep{lin_evolution_1985,audard_episodic_2014}. Several theoretical models have tried to explain these episodic accretion events \citep{bell_using_1994,armitage_episodic_2001,zhu_nonsteady_2009,faure_thermodynamics_2014}. One explanation on which we focused is the thermal instability (TI) at the inner edge of the dead zone, where models predict an increase in the disc turbulence with rising temperatures (around 1000K; \citealt{williams_ionization_2024}) because the gas is ionised enough to trigger magneto-rotational instability (MRI). 

%explain the burst cycle
The burst cycle triggered by TI, as predicted by theoretical models \citep{ziampras_planet_2026,cecil_variability_2024,chambers_large_2024,cecil_mri-triggered_2026}, usually evolves as follows: in the quiescent phase, gas and dust accumulate at the inner edge of the dead zone due to the local radial gradient in turbulence. When the material at the inner edge of the dead zone reaches the MRI activation temperature because viscous heating overcomes the radiative cooling, a runaway heating phase is triggered that in turn increases the turbulent viscosity. The increased viscosity partially drives material outwards, igniting the MRI along the way until the surface density and associated viscous heating are too low to sustain the burst front. Simultaneously, the increased viscosity flushes more material onto the star, increasing its accretion rate. As the disc cools again, several smaller bursts are re-triggered at the inner edge of the dead zone in some models, \citep[e.g.][]{ziampras_planet_2026,cecil_variability_2024}, whereas other models return directly to the quiescent phase \citep[e.g.][]{cecil_mri-triggered_2026}. These re-triggered bursts propagate progressively shorter distances until the disc returns to the quiescent phase.

These periodic TI-triggered accretion outbursts have been studied in multiple previous works. While early works relied on 1D models with simplified heating and cooling \citep[e.g.][]{armitage_episodic_2001}, more recent studies have included more sophisticated heating and cooling \citep{chambers_large_2024,owen_importance_2014}, as well as dust coagulation and evolution and a proper coupling of dust properties and opacities \citep{ziampras_planet_2026-1,kato_thermal_2026}. The problem was also studied in non-vertically integrated radiation hydrodynamic simulations \citep[e.g.][]{wunsch_two-dimensional_2006,cecil_variability_2024} that found that the burst can lead to several smaller re-flares, leading to the formation of multiple rings in the post-burst state. However, the non-axisymmetric simulations by \citet{ziampras_planet_2026} revealed that the burst front becomes Rossby-wave unstable \citep{lovelace_rossby_1999}, leading to the formation of vortices that smooth the rings out. One part of physics that has been largely neglected is that the dust evaporates during the outburst. This was only considered in \cite{ziampras_planet_2026-1} and \citet{cecil_variability_2024}, who reduced the dust opacity as a function of temperature \citep{isella_shape_2005}. However, it is important to consider this effect because the evaporation and condensation of dust during the outburst changes its dynamical coupling to the gas, which changes the redistribution of the dust during the burst and leads to a different post-burst state.

On the other hand, several studies have investigated the composition of the inner disc using coupled chemical and dust evolution models. These models either considered only the evaporation and condensation of chemical components  \citep[e.g.][]{mah_mind_2024,williams_locked_2025} or included chemical networks with varying levels of complexity \citep[e.g.][]{booth_planet-forming_2019,sellek_chemical_2025,bosman_efficiency_2018,molyarova_metamorphoses_2026}. The addition of chemical reactions has been shown to modify the bulk properties of the disc compositions on timescales of $10^5-10^6$ yr \citep{eistrup_molecular_2018,bosman_efficiency_2018}. For example, the C/O ratio in the inner disc has been found to be altered by chemical reactions and the associated change in radial transport for $t > 1,\rm Myr$ \citep{molyarova_metamorphoses_2026,sellek_chemical_2025}, as these reactions change the dominant chemical components delivering carbon to the inner disc. Nevertheless, it remains essential to accurately model the dust dynamics, including evaporation and condensation, because the drift timescales are typically shorter than the timescales on which chemical reactions affect the composition for high turbulence ($\alpha > 10^{-3}$) \citep{booth_planet-forming_2019}. These studies represent another step towards interpreting the composition of the inner disc, which has been the focus of multiple observational programmes with James Webb Space Telescope (JWST) \citep{henning_minds_2024,kamp_chemical_2023}. Thus far, previous simulations mainly focused on the compositional imprint of the dust transport from the outer disc and did not consider the outbursts in the inner disc that can reset the inner disc chemistry.

However, previous studies have not combined modelling of the inner disc composition with the aforementioned outbursts. Since outbursts might have long-lasting imprints on the  inner disc chemistry \citep{houge_burned_2025}, studying the two concurrently might yield new insights into the chemical evolution of the inner disc. 
Therefore, we investigated the effect of dust evaporation and condensation on the dynamics of the outburst. We did this by running vertically integrated axis-symmetric models that included dust coagulation and evolution using the \texttt{TriPoD} method and included evaporation and condensation of dust calculated consistently with its opacity-dependent temperature. This allowed us to track the disc composition by considering a condensation sequence of chemical components that make up the dust and its respective dust vapour. This allowed us to model the effect of these outbursts  on the inner disc composition, and vice versa.

This work is structured as follows: in Sect.~\ref{sec:model} we describe the model we used for the simulations, and in Sect.~\ref{sec:results}, we present the results of our different setups testing the effect of dust sublimation on the outbursts and the simulations with a full composition. This is followed by a discussion of our results in Sect.~\ref{sec:discussion}. Lastly, we provide a final overview and outlook in Sect.~\ref{sec:conclusion}.

\section{Model description}
\label{sec:model}
In this section, we describe the model we used to simulate the accretion outbursts in our simulation. Additionally, we present the initial conditions and the setup we used to generate our results.

\subsection{Gas and dust evolution}
The disc was simulated in 1D, assuming azimuthal symmetry, where the vertical gas structure was given by hydrostatic equilibrium. The evolution of the gas disc was simulated using the canonical viscous $\alpha$-description \citep{shakura_black_1973}, that is, the disc gas surface density $\Sigma_{\rm g}$ was evolved according to the following equation:
\begin{equation}
\frac{\partial}{\partial t} \Sigma_{\rm g} =  \frac{1}{R} \frac{\partial}{\partial R} \left[3R^{1/2} \frac{\partial}{\partial R} (R^{1/2} \Sigma_{\rm g} \nu_{\rm g}) \right], 
\end{equation}
with $\nu_g = \alpha c_s^2/\OmegaK$ being the viscosity, and the sound speed is given by $c_s = \sqrt{k_b T/\mu}$ as a function of temperature $T$, the is mean molecular weight $\mu$, and $\OmegaK$ is the Keplerian angular velocity. 

The dust was modelled using the \tripod{} method \citep{pfeil_tripod_2024}, which describes the dust size distribution of spherical grains as a truncated power law $n(a)\propto a^q \Rightarrow \sigma(a) \propto a^{q+3}$, where $a$ is the dust size, and $\sigma(a)$ is the surface density of a specific grain size such that $\int_0^\infty \sigma(a) da = \Sigma_{\rm d}$. We used the \tripodpy implementation of the algorithm \citep{kaufmann_tripodpy_2025} that tracks the compositional evolution of multiple gas and dust components through transport, evaporation, and condensation. For details of the \tripod method, we refer to \cite{pfeil_tripod_2024} and Appendix \ref{ap:tripod} and only give a brief overview here.

The dust size distribution is parametrised by three variables: the surface density of the small and large grains $\Sigma_0$ and $\Sigma_1$, and the maximum grain size \amax, where the minimum grain size $a_{\rm min}$ is a constant. The surface densities $\Sigma_0$ and $\Sigma_1$ describe the mass in particles between $[a_{\rm min}, a_{\rm int}]$ and  $[a_{\rm int}, a_{\rm max}]$, respectively, where $a_{\rm int} = \sqrt{a_{\rm max} a_{\rm min}}$. The power-law exponent of the size distribution can be reconstructed from the parameters via
\begin{equation}
    q  = 2\frac{\log(\Sigma_1/\Sigma_0)}{\log(a_{\rm max}/a_{\rm min})} -4.
\end{equation}
The radial drift of the two surface densities was calculated with their mass-averaged particle sizes, which can be calculated by integrating over the size distribution. Additionally, the two mass bins exchanged mass on their collisional timescales, relaxing to the appropriate dust size distribution power law expected from theory \citep{birnstiel_dust_2011}. The expected power law depends on whether the dust is in the fragmentation limit with $q_\mathrm{frag} \in [-3.5,-3.75]$ or in the drift limit and the growth phase, which leads to $q_\mathrm{grow} = -3$ \citep{birnstiel_dust_2024}.

The growth of the maximum grain size of the size distribution was modelled using a mono-disperse approximation, but using the relative velocity between grains of different sizes to infer the growth rate and adding a transition function to stop growth at the fragmentation barrier. Additionally, \amax~was advected as a passive scalar that follows the radial transport of $\Sigma_1$.

\subsection{Composition tracking}
When considering compositional tracking of different molecules in the gas and dust, we split up the total gas and dust surface density into the sum of its components, that is, $\Sigma_{\rm g/d} = \sum_i \Sigma_{{\rm g/d}, i}$. The evolution of the total gas can be described by the sum of the evolution of its individual components (i.e. the different elements and molecules) by solving \citep{clarke_diffusion_1988,morfill_transport_1984, pavlyuchenkov_dust_2007}
\begin{equation}
\label{eq:evol_gas_compo}
\frac{\partial \Sigma_{\text{g},i}}{\partial t}
+ \frac{1}{R} \frac{\partial}{\partial R} (R \Sigma_{\text{g},i} v_R)
= \frac{3}{R} \frac{\partial}{\partial R} \left[ R \nu \Sigma_{\rm g} \frac{\partial}{\partial R} \left( \frac{\Sigma_{\text{g},i}}{\Sigma_{\rm g}} \right) \right] + \sum_j \left. \frac{\partial \Sigma_{\text{g},i,j}}{\partial t}\right\vert_{\rm sub}.
\end{equation}
The gas velocity $v_R$ is given by
\begin{equation}
\label{eq:vr_compo}
v_R = -\frac{3}{\Sigma_{\rm g} \sqrt{R}} \frac{\partial}{\partial R} \left( \Sigma_{\rm g} \nu_{\rm g} \sqrt{R} \right).
\end{equation}
This decomposition assumes a ratio of viscosity and diffusivity (Schmidt number) of  $\text{Sc}_{\rm g} = \nu_{\rm g}/D= 1/3$ \citep[following][]{pavlyuchenkov_dust_2007}, which may not be valid in general (for a discussion, see \citealt{houge_burned_2025}). Since the different components of the gas have different mean molecular weights, we calculated the total mean molecular weight as
\begin{equation}
    \mu = \frac{\Sigma_{\rm g}}{\sum_i \Sigma_{\rm g,i}/\mu_i}.
\end{equation}

To track the different dust components, we followed the approach of \cite{booth_planet-forming_2019} and separated the evolution of the dust into its components as
\begin{align}
\label{eq:evol_dust_compo}
& \frac{\partial \Sigma_{\text{d},i,j}}{\partial t} + \frac{1}{R} \frac{\partial}{\partial R} \left\{ R \left[ \Sigma_{\text{d},i,j} \cdot u_{\text{d},j} - \frac{\partial}{\partial R} \left( \frac{\Sigma_{\text{d},i,j}}{\Sigma_{\text{gas}}} \right) \cdot \Sigma_{\text{gas}} D_j \right] \right\} \nonumber \\
& =  \left. \frac{\partial \Sigma_{\text{d},i,j}}{\partial t}\right\vert_{\rm coag} - \left. \frac{\partial \Sigma_{\text{d},i,j}}{\partial t}\right\vert_{\rm sub},
\end{align}
where the index $i$ refers to the individual component, and $j$ refers to the mass bin (i.e. either the small or large grains). The dust diffusivity $D_j = \frac{\alpha c_s^2}{\Omega_K (1 + \rm St_j^2)}$ (this is a standard choice in the literature \citep[e.g.][]{stammler_redistribution_2017,stammler_dustpy_2022,schneider_how_2021}, even though it is inconsistent with the gas diffusivity because it implies $\rm Sc = 1$
) and radial velocity $u_{d,j}$ calculated as in \cite{stammler_dustpy_2022} were computed using the mass-weighted average size in each bin, that is, $a_0 = \langle a\rangle_{[a_{\rm min},a_{\rm int}]}$ and $a_1 = \langle a\rangle_{[a_{\rm int},a_{\rm max}]}$. The coagulation source term for each component was calculated using the same timescale as for the global dust, that is,
\begin{align}
\left. \dot{\Sigma}_{\text{d},i,0}\right \vert_{\rm coag} &= - \left. \dot{\Sigma}_{\text{d},i,1}\right \vert_{\rm coag} = \dot{\Sigma}_{i,1\rightarrow 0} - \dot{\Sigma}_{i,0 \rightarrow 1}\\ 
\dot{\Sigma}_{\text{d},i,0 \rightarrow 1} &= \frac{\Sigma_{\text{d},i,0} \Sigma_{\text{d},1} \sigma_{01} \Delta v_{01}}{m_1 \sqrt{2\pi(H_0^2 + H_1^2)}},
\dot{\Sigma}_{\text{d},1\rightarrow 0}
= \frac{\Sigma_{\text{d},1,i}\Sigma_{\text{d},1}\,\sigma_{11}\,\Delta v_{11}}{ m_1\,2\pi H_1^{2}}\,\mathcal{F},
\end{align}
where $\sigma_{i,j} = \pi(a_i+a_j)^2$ is the geometrical cross section, and the relative velocities in the collisions were calculated as in \cite{stammler_dustpy_2022} for the sizes $\Delta v_{11} = \Delta v(a_1,0.4a_1)$ and $\Delta v_{01} = \Delta v(a_0,a_1)$. The size distribution calibration function $\mathcal{F}$ ensures that the mass exchange leads to the appropriate equilibrium dust size power law as defined in \eqref{eq:F_mathcal}. The size distribution of each component relaxes to the expected power law on the collisional timescale of the total dust. As the different components can have different bulk densities ($\rho_{s,i}$), the total bulk density of the dust is given by
\begin{equation*}
\rho_s = \sum_i \Sigma_{\text{d},i} /(\sum_i \Sigma_{\text{d},i} /\rho_{s,i}).
\end{equation*}

To model the evaporation and condensation of the individual components, we added sources and sink terms between the dust components and their counterpart in the gas for evaporation and condensation \citep{cuppen_grain_2017,penzlin_bowie-align_2024}. The sublimation rate of each component can be described by
\begin{equation*}
\dot{\Sigma}_{S,i,j} =
\frac{\Sigma_{\text{d},j} \langle 4\pi a^{2} \rangle_j}{\langle m_d \rangle_j}
N_{\mathrm{bind}}\, m_i\, f_{C,i,j}\, \nu_i
\exp\!\left(-\frac{E_{\mathrm{sub},i}}{T}\right),
\end{equation*}

where $f_{C,i,j}$ is the fraction of binding sites covered by the chemical component $i$ on the grains of mass bin $j$. When the filling of each binding site is random, $f_{\rm C,i,j}$ is given by \citep{robinson_co_2026}
\begin{equation*}
f_{C,i,j} = 1 - \text{exp}\left(-\frac{\Sigma_{d,i,j}}{\frac{\Sigma_{\text{d},j} \langle 4\pi a^{2} \rangle_j}{\langle m_d \rangle_j}N_{\mathrm{bind}} \:m_i}\right).
\end{equation*}
The argument of the exponent is the ratio of the number of particles of component $i$ in mass bin $j$ divided by the total number of binding sites available on the surface of the grains. The formula transitions smoothly from $f_{C,i,j}=1$ for fully covered grains to the first-order rate for partial coverage.
The total surface area of the grains in each mass bin $\frac{\Sigma_{\text{d},j}\langle 4\pi a^{2} \rangle_j}{\langle m_d \rangle_j}$ was obtained by analytically integrating over the size distribution. The counteracting sink term describing the condensation is given by
\begin{equation*}
\dot{\Sigma}_{C,i,j} =
\frac{\Sigma_{\text{d},j}\langle 4\pi a^{2} \rangle_j}{\langle m_d \rangle_j}
\frac{\Sigma_{g,i}}{\sqrt{2\pi}\,H}
\sqrt{\frac{8 k_B T}{\pi m_i}}\, P_{\mathrm{stick}}.
\end{equation*}

The quantities $\nu_i$ (trial frequency), $m_i$ (molecular weight), and $E_{sub,i}$ (binding energy expressed in units of Kelvin) are material-dependent quantities that determine the thermodynamic properties of each component. Furthermore, we assumed a sticking probability $P_{\rm stick} =1$ and a number of binding sites per surface area of $N_{\rm bind} = \SI{1.5e15}{cm^{-2}}$ for this work. Combined, these two terms give the sublimation condensation source term,
\begin{equation}
\left. \frac{\partial \Sigma_{\text{d},i,j}}{\partial t}\right\vert_{\rm sub} = \dot{\Sigma}_{S,i,j} - \dot{\Sigma}_{C,i,j}.
\end{equation}
The growth rate and advection of the maximum particle size were still computed using the total dust surface density; this means that different dust components all have the same maximum grain size, but can be distributed unevenly among the different grain sizes, that is, 
\begin{equation}
    q = \frac{\text{log}(\Sigma_{\text{d},1}/\Sigma_{\text{d},0})}{\text{log}(\sqrt{a_{\rm max}/a_{\rm min})}} -4 \neq q_i = \frac{\text{log}(\Sigma_{\text{d},i,1}/\Sigma_{\text{d},i,0})}{\text{log}(\sqrt{a_{\rm max}/a_{\rm min})}} -4.
\end{equation}

A comparison of our condensation and evaporation module with the more detailed Smoluchowski solver \cudisc \citep{robinson_introducing_2024,robinson_co_2026}, which also includes compositional tracking, can be found in Appendix \ref{ap:cudisc}

\subsection{Temperature evolution}
The temperature evolution of the disc was modelled by considering viscous heating ($Q_{\rm visc}$), stellar irradiation ($Q_{\rm  irr}$), and radiative cooling ($Q_{\rm  cool}$), which are given by the following terms \citep{ziampras_spirals_2025}:
\begin{subequations}
\label{eq:Q}
\begin{align}
\label{eq:Q_visc}
Q_{\rm visc} &= \frac{9}{4} \alpha c_s^2\Sigma_g\Omega_K \\
\label{eq:Q_irr}
Q_{\rm irr} &=  2 \frac{L_\star}{4\pi r^2} (1 -\varepsilon) \frac{\theta}{\tau_{\text{eff}}} \\
\label{eq:Q_cool}
Q_{\rm cool} &= -2 \frac{\sigma_{SB}T^4}{\tau_{\text{eff}}},
\end{align}
\end{subequations}
where $\tau_{\rm eff}$ is the effective optical depth, $\theta$ is the grazing angle, and $\varepsilon = 0.5$ is the disc albedo. The resulting change in temperature can be defined via the internal energy as
\begin{equation}
    \frac{dT}{dt} = \frac{\mu*(\gamma-1)}{\Sigma_g k_B} \times [Q_{\rm visc} + Q_{\rm visc} + Q_{\rm cool}],
\end{equation}
with $\gamma = 7/5$.
 Additionally, temperature was treated as a passive scalar that is advected and diffused along with the gas. The effective optical depth is defined by the Planck and Rosseland opacity via \citep{hubeny_vertical_1990},
\begin{equation}
    \tau_{eff} = \frac{3\tau_{\rm R}}{8} + \frac{\sqrt{3}}{4} + \frac{1}{4\tau_{\rm P}},
\end{equation}
where $\tau_{\rm R,P} = \frac{\kappa_{\rm R,P}\Sigma_g}{2}$, and the Planck and Rosseland opacities were calculated from the dust ($\kappa_{\rm d}$) and gas opacities ($\kappa_{\rm g}$),
\begin{equation}
    \kappa = \epsilon \kappa_{\rm d} +\kappa_{\rm g}.
\end{equation}
We took the gas opacity to be constant, with $\kappa_{\text{R,P},g} = 10^{-3}\rm cm^2/g$. The dust opacities were calculated using the \texttt{growpacity} package \citep{ziampras_growpacity_2026} using tabulated results obtained with \texttt{optool} \citep{dominik_optool_2021} for different temperatures and dust size distributions (dust power law $q$ and maximum grain size $a_{\rm max}$), and we calculated the dust opacities via interpolation during runtime. We used the default DIANA composition \citep{woitke_consistent_2016} for \texttt{optool} to calculate the tabulated grid.

The heating and cooling rates above can lead to a thermal instability when the heating rate is more strongly dependent on temperature than the cooling rate, that is,
\begin{equation}
    \label{eq:stability_crit}
    \frac{\partial Q_{\rm heat}}{\partial T}  > \frac{\partial \vert Q_{\rm cool}\vert}{\partial T}.
\end{equation}
 This can occur either when the gas opacities transition at high temperature or at the edge of the dead zone \citep{armitage_protoplanetary_2019}, when the turbulence is a strong function of temperature because the MRI is activated, which we investigated here.

\subsection{Treatment of the inner edge}
To model the inner disc edge, we took a few additional effects into account that are listed below. 
Firstly, at high temperatures, we accounted for the higher turbulence that is triggered by the MRI. To model this, we followed the approach of \citep{cecil_variability_2024} by considering a temperature-dependent $\alpha-$parameter,
\begin{equation}
    \label{eq:alpha_mri}
    \alpha (T) = \alpha_0 + \frac{1}{2} (\alpha_{\rm MRI} - \alpha_0) \left[ 1 + \text{tanh}\left(\frac{T-T_{\rm MRI}}{\Delta T_{\rm MRI}}\right)\right],
\end{equation}
where $\alpha_0 = 10^{-3}$ is the dead-zone value, $\alpha_{\rm MRI} = 0.1$, $T_{\rm MRI}=900\,\rm K$, and $\Delta T_{\rm MRI} = 25\,\rm K$.
Secondly, to account for the direct irradiation from the star, we modified the grazing angle in Eq. \eqref{eq:Q_irr} from $\theta_0 \approx \frac{8H}{7R}$ \citep{dullemond_are_2000} to
\begin{equation}
    \label{eq:theta_irr}
    \theta(r) = \theta_0 + \frac{1}{2} (1-\theta_0)\left[1-\tanh{\left(\frac{r-r_{\rm rim}}{0.01 \rm au}\right)}\right]
\end{equation}
to account for the direct irradiation from the star at the inner rim \citep{ziampras_planet_2026-1}, where we set $r_{\rm rim} = 0.1 \rm$ au.

\subsection{Dust composition}
\label{sec:dust_comp}
For the dust composition, we had two different setups. The first setup modelled the dust as a single component with a single sublimation temperature, and the second setup consisted of a condensation sequence that represents the silicates through six main components. Although the first approach is unrealistic, it let us isolate the effect of the dust sublimation temperature on the evolution of the accretion bursts. 

For the single-component models, we chose different sublimation energies to study the following cases: dust with an extremely high binding energy ($E_{\rm sub} = \SI{5e4}{K}$), dust sublimating at the hottest sublimation temperature of corundum ($E_{\rm sub} = \SI{4.2e4}{K}$), dust sublimating at the sublimation line of the majority of dust near fosterite and enstatite ($E_{\rm sub} = \SI{3.5e4}{K}$), dust sublimating just above the MRI-activation temperature ($E_{\rm sub} = \SI{3.1e4}{K}$), dust sublimating below the MRI-activation temperature ($E_{\rm sub} = \SI{2.6e4}{K}$), and dust matching the soot line ($E_{\rm sub} = \SI{2.1e4}{K}$). All model setups and different values we explored for the single dust component approach are listed in Table \ref{tab:dust-compo}. 

The second setup considered dust consisting of different minerals with different binding energies, mean molecular weights, and bulk densities. For the sequence of condensing minerals, we included corundum as Al$_2$O$_3$, hibonite as CaAl$_{12}$O$_{16}$, melilite as Ca$_2$Al$_2$SiO$_7$, pyroxene as CaMgSi$_2$O$_6$, fosterite as Mg$_2$SiO$_4$, and enstatite as MgSiO$_3$. As we only considered sublimation and condensation chemistry, we did not include reactions of fosterite to enstatite or those of hibonite to melilite. 
To approximate these processes, we adjusted the abundances of these materials to ensure that the dust had the correct mass fraction at the different condensation lines. We fitted the sublimation energies and fractional abundances between components to the values in \cite{yoneda_condensation_1995}. We chose the fractional abundance of the mineral species relative to either the protosolar (Al/H)$_*$ abundance of $10^{-5.85}$ or the protosolar (Mg/H)$_*$ abundances of $10^{-4.7}$, with the fractions given in Table \ref{tab:dust-compo}, to represent the full reservoir of dust beyond the sublimation lines. Corundum, hibonite, and melilite account for the total sluminium, while pyroxene, fosterite, and enstatite account for the total magnesium and silicates in the disc. To limit the number of species, we did not include iron, which has a similar sublimation temperature as enstatite and fosterite and is generally less abundant then these.
The minerals take up about 20.9\% of the total oxygen budget. 

In addition, we included four volatiles with sublimation lines in the inner disc: soot, which is tracked as C$_4$H$_{10}$ in the gas phase (also referred to as C-grains), water, methanol, and carbon dioxide. 
Methanol was chosen to represent the intermediately volatile organic species that might be affected by the strong burst reported in \cite{ziampras_planet_2026}.
The abundance of these sums up to two-thirds of the remaining oxygen budget because the remaining oxygen might be bound in CO, metal oxides, or further organic species. We further chose the C-grains to be as abundant as half of the water. This is an arbitrary choice because the real amount of solid carbon material relative to other more volatile carbon species is highly uncertain.

\begin{table*}[]
    \centering
    \caption{Material-dependent parameters of the dust components for the condensation and single-dust models.}
    \label{tab:dust-compo}
    \begin{tabular}{l c c c c c c c}
        Name& Formula &Abundance & Mass fraction $[\Sigma_{\rm g}]$ & $\mu$ $[\mathrm{m_p}]$& $\rho_\text{bulk}$ $[\mathrm{g/cm^3}]$& $E_\text{sub}$  $[10^3 \mathrm{K}]$& $\nu_i$\\
        \hline
        \hline
        Corundum & (Al$_2$O$_3$) & 0.5/2 (Al/H)$_*$ &$6.98\times10^{-5}$ &100 & 4.02& 44.25& $10^{12}$\\
        \hline
        Hibonite  & (CaAl$_{12}$O$_{16}$) & 0.2/12 (Al/H)$_*$ & $3.05\times10^{-5}$ & 656 & 3.84& 43.575& $10^{12}$\\
        \hline
        Melilite  & (Ca$_2$Al$_2$SiO$_7$) & 0.3/2 (Al/H)$_*$ & $1.13\times10^{-4}$ & 272 & 2.95& 40.7& $10^{12}$\\
        \hline
        Pyroxene  & (CaMgSi$_2$O$_6$) & 1/28 (Mg/H)$_*$ & $3.04\times10^{-4}$ & 216 & 3.4& 36.225& $5\times10^{12}$\\
        \hline
        Fosterite  & (Mg$_2$SiO$_4$) & 19/56 (Mg/H)$_*$& $1.87\times10^{-3}$ &140 & 3.27& 36.075& $5\times 10^{12}$\\
        \hline
        Enstatite & (MgSiO$_3$) & 2/7 (Mg/H)$_*$ & $1.12\times10^{-3}$ & 100& 3.2& 34.15& $5\times10^{12}$\\
        \hline
        Carbon-chains& (C$_n$H$_m$) & 1/6 (O/H)$_\text{gas}$ & $3.75\times10^{-3}$ & 58 & 2.27 & 19.05& $4\times 10^{13}$\\        \hline
        Water & (H$_2$O) & 1/3 (O/H)$_\text{gas}$ &  $2.23\times10^{-3}$ & 18 & 1.0 & 5.8& $4 \times10^{13}$\\
        \hline
        Methanol & (CH$_3$OH) & 1/6 (O/H)$_\text{gas}$ & $2.07\times10^{-3}$ & 20 & 1.7 & 4.93& $6\times 10^{16}$\\
        \hline
        Carbon dioxide & (CO$_2$) & 1/12 (O/H)$_\text{gas}$ &  $1.42\times10^{-3}$ & 44 & 1.56 & 2.7& $10^{13}$\\
        \hline
        \hline
        one-component& & & 0.01 & 87 & 1.67 & \makecell{[50,42,35,\\31,26,21]} & $1.4\times10^{14}$\\

    \end{tabular}
\end{table*}

\begin{figure}
    \centering
    \includegraphics[width=\linewidth]{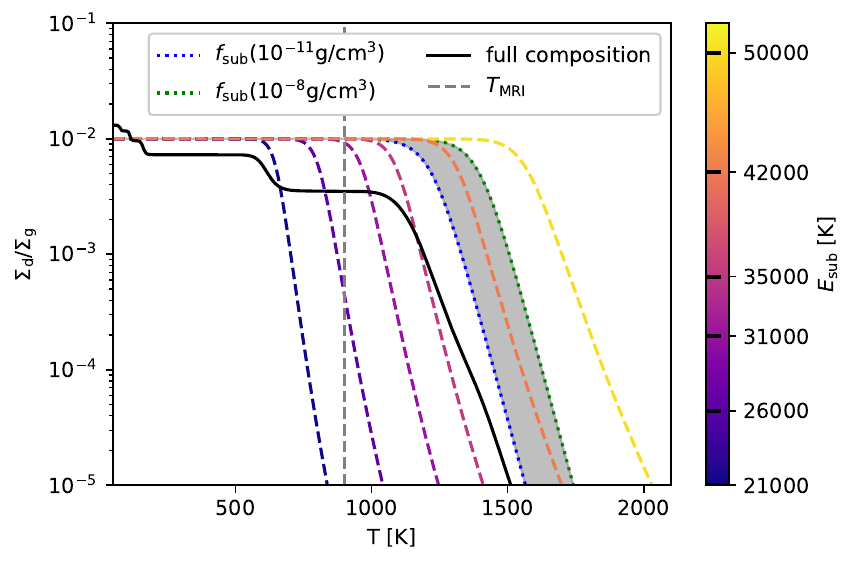}
    \caption{Initial dust-to-gas ratio as a function of temperature for our different setups, i.e. single-component setup (dashed), full composition (solid), and scalar treatment (shaded region and dotted lines) }
    \label{fig:eps_T_setups}
\end{figure}

Lastly, to compare setups with the works from \cite{ziampras_planet_2026-1}, we compared our sublimation curves with the non-sublimating dust where the opacities are reduced at high temperatures to mimic sublimation according to
\begin{align}
    \kappa_{\rm d,P/R} &= \kappa_0 * f_{\rm sub}\\
    f_{\rm sub} &= 10^{-10} + \frac{1}{2}(1-10^{-10}) \left[ 1 - \text{tanh}\left( \frac{T-T_{\rm sub}}{100 \rm K}\right)\right] \\
    T_{\rm sub} &= 2000 \cdot \left(\frac{\rho_{\rm g}}{\rm g/cm^3}\right)^{0.0195} K,
\end{align}
where $\kappa_0$ is the opacity calculated as described above, $\rho_{\rm g}$ is the mid-plane gas density, and $T_{\rm sub}$ was defined in \cite{isella_shape_2005}. To show the resulting condensation curves in the different setups, we plot the initial dust-to-gas ratio as a function of temperature in Fig.~\ref{fig:eps_T_setups}.

\subsection{Initial conditions}
A few simulation properties were kept constant in all our simulations. Firstly, the host star in all simulations had a mass of  $M_{\star} = 1\,\Msun$ and a luminosity of $L_\star = 1.78\,\Lsun$, with a radius of $R_\star = 2.5\,\Rsun$. The radial grid of the simulations extended from $R_{\min} = 0.05 \: \rm au$ to $R_{\max} = 5 \:\rm au$, with $N_r = 1000$ logarithmically spaced grid cells. The dust was initialised with an MRN-like distribution \citep{mathis_size_1977} (i.e. $q = -3.5$ and $a_{\max} = 10^{-4}$ cm) and the fragmentation velocity was kept at $v_{\rm frag} =1 \rm m/s$. The initial surface density was chosen to be in an unstable state given by
\begin{equation}
    \Sigma_g = 200 (\text{max}\{R_{\rm max},R\}/{\rm au})^{-15/14}\rm{g/cm}^2,
\end{equation}
where $R_{\rm max}$ was chosen so the initial cleared the disc until this radius at least, flushing the parts of the disc interior to it. The first burst cycle was then discarded as its evolution is governed by the initial conditions. The mean molecular weight of the uncontaminated background gas is given by $\mu =2.3\,m_p$. The dust surface density for the single-component case is given by $\Sigma_{\rm d,i} + \Sigma_{\rm g,i} = 0.01 \Sigma_{\rm g}$ as a constant fraction of the background gas. When we considered evaporating dust, we initialised it in the sublimation evaporation equilibrium with its respective vapour to be consistent with the initial temperature structure, that is, $\left. \frac{\partial \Sigma_{\text{d},i,j}}{\partial t}\right\vert_{\rm sub} = 0$. When we considered the full condensation sequence, the dust was separated into the different components according to abundances in Table \ref{tab:dust-compo}, which were chosen so that we recovered a total dust to gas ratio of $\Sigma_{\rm d} = 0.01 \Sigma_{\rm g}$ beyond the water snow line, as shown in Fig. \ref{fig:eps_T_setups}. We initialised the temperature in the equilibrium between heating and cooling $Q_{\rm irr}(T_0) = - Q_{\rm cool}(T_0)$. We chose the dead zone $\alpha$ as $\alpha_{\rm 0} = 10^{-3}$. For the outer boundary condition, we enforced a steady-state gas profile with $\Sigma_{\rm gas} \propto R^{-15/14}$ and a dust-to-gas ratio of 1\%.

\section{Results} 
\label{sec:results}
In this section, we present the results for our different setups. First, we describe in Section \ref{sec:single_comp} the single-component simulations and the effect of the sublimation temperature on the burst behaviour, and we then investigate the effect of changing other disc parameters on our results in Section \ref{sec:disk_params}. Lastly, we study the setups using the dust with the full composition in Section \ref{sec:full_comp}.
\subsection{Single-component dust}
\label{sec:single_comp}
To assess the effect of the dust sublimation temperature on the burst dynamics, we compared the outcomes of single-component simulations with binding energies ranging from $E_{\rm sub} = 21$'000 to 50'000 K, as listed in Table \ref{tab:dust-compo}.

\begin{figure*}
     \centering
     \includegraphics[width=\linewidth]{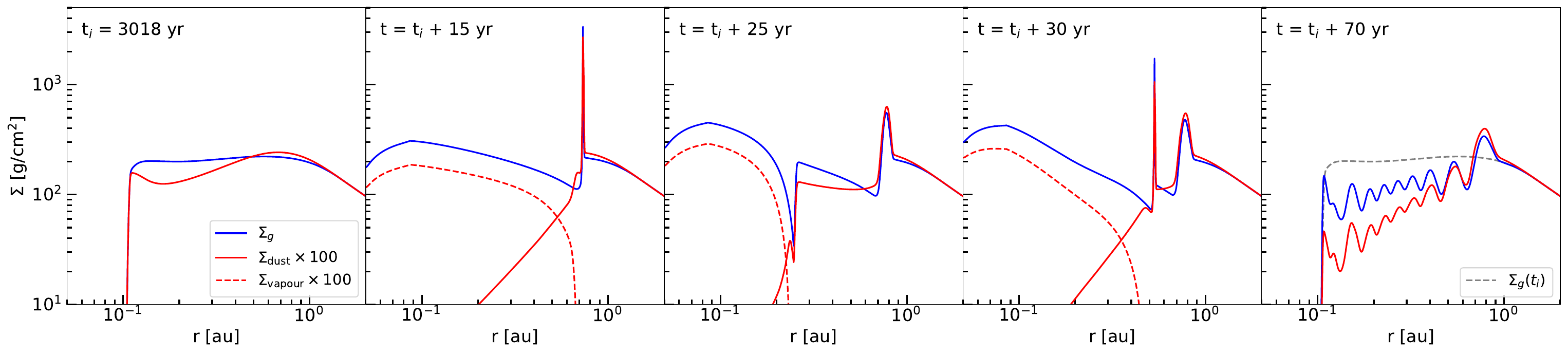}
     \caption{Gas (blue) and dust, and the associated vapour (red) surface density during a burst for $E_{\rm sub} = 4.2\times10^4 \rm K$}
     \label{fig:cycle_nom}
 \end{figure*}

\begin{figure*}
    \includegraphics[width=\linewidth]{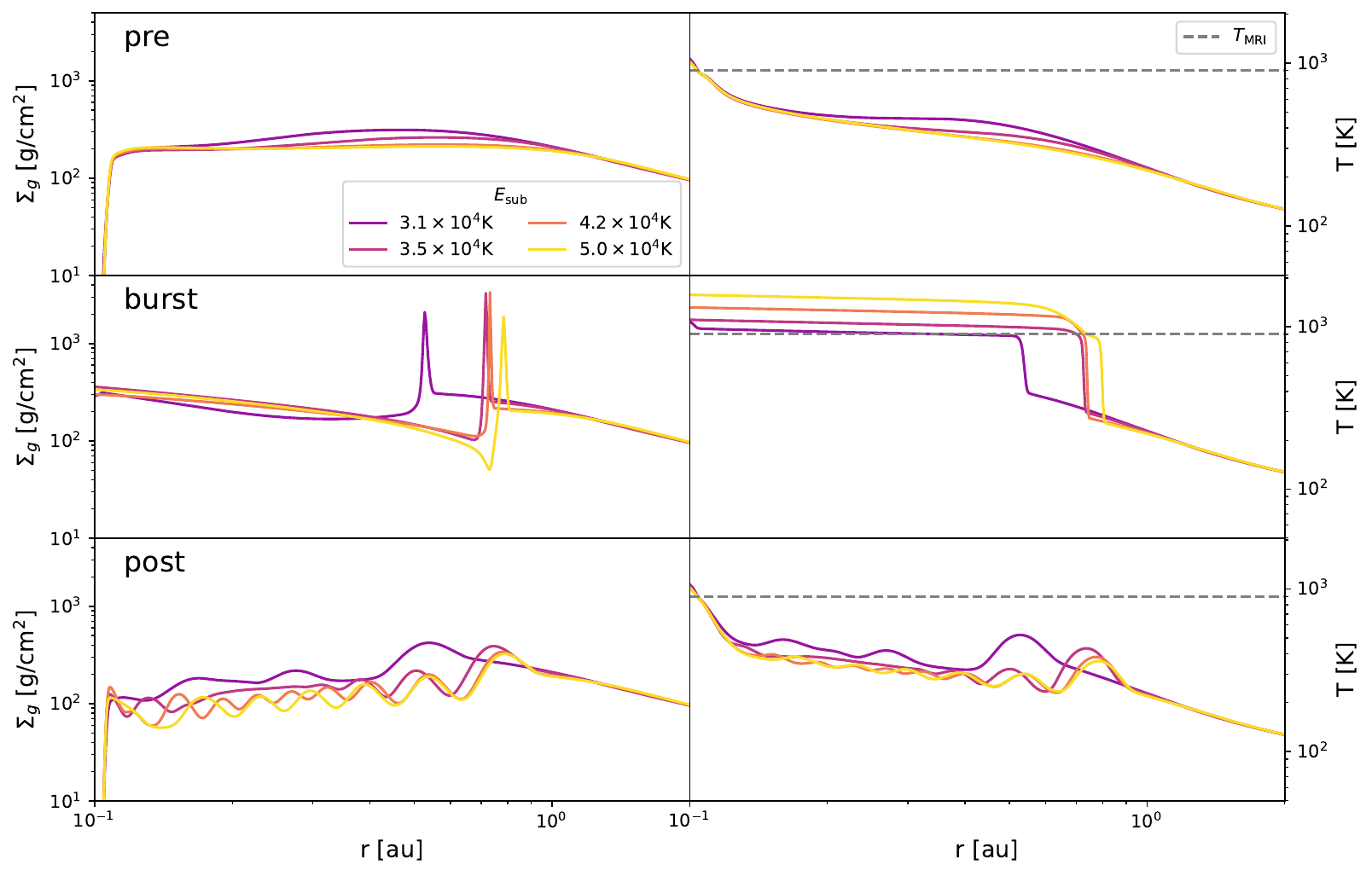}
    \caption{States of the single-component setups before, during, and after a burst, where we show the gas surface density(left) and the temperature (right)}
    \label{fig:burst_states}
\end{figure*}

 Focusing on the nominal case ($E_{\rm sub} = \SI{4.2e4}{K}$), we observed the same general outburst behaviour as described in \cite{cecil_variability_2024} and \cite{ziampras_planet_2026-1}, which is shown in Fig. \ref{fig:cycle_nom}. When the inner disc edge had accumulated enough material from viscous evolution, the viscous heating was strong enough to push the temperature to the MRI activation temperature (panel (a), and the temperature is shown in the top panel of Fig. \ref{fig:burst_states}), where it triggered the runaway heating to the high-temperature equilibrium given by the dust sublimation. The contrast in turbulence caused material to diffuse outwards, heating the disc up to $\sim 0.9$ au (panel (b)). As the wave faded, the in-flushing material triggered several re-flares (panels (c) and (d)), launching waves that travel shorter distances each time, leaving behind multiple rings where the gas piled up, as seen in the post-burst state, before returning to the quiescent state with a significantly depleted disc in the outbursting region (panel (e)). The cycle then repeated when viscous evolution refilled the burst region to the pre-burst state.

\begin{figure}
    \centering
    \includegraphics[width=\linewidth]{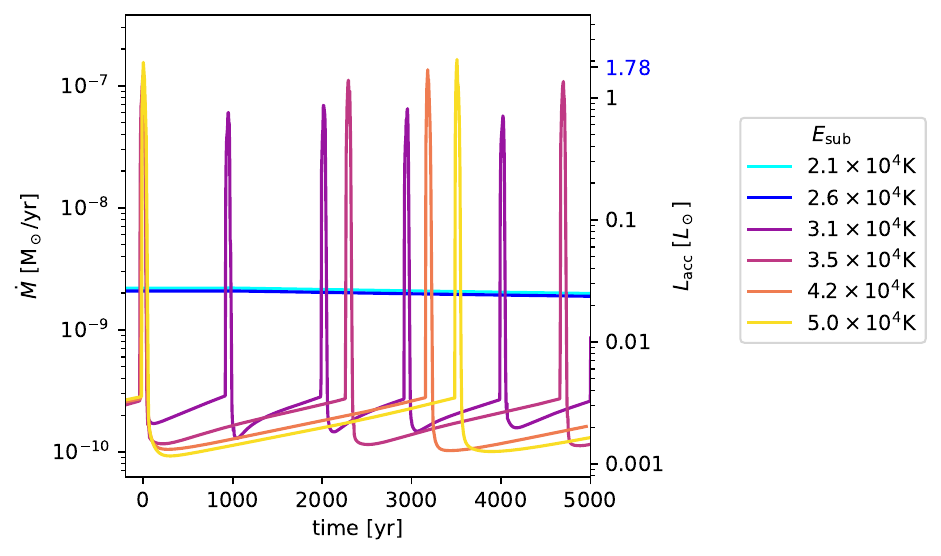}
    \caption{Accretion rate/luminosity through the inner boundary for the different binding energies as a function of time}
    \label{fig:acc_rate}
\end{figure}

Next, we compared all the simulations that were able to trigger an accretion outburst during the runtime, which were all the simulations with binding energies $E_{\rm sub} > \SI{26 000}{K}$. In Fig. \ref{fig:burst_states} we show the state of these simulations before, during, and after the burst, namely their gas surface density and temperature. Multiple clear trends in the burst behaviour scale with the dust sublimation temperature. Firstly, the distance traveled by the burst increases with increasing sublimation temperature and saturates for $E_{\rm sub} = \SI{4.2e4}{K}$ and $\SI{5e4}{K}$ around 0.9 au, as seen in the outermost surface density bump of the post-burst states. The second clear trend is that the temperature in the hot state increases with increasing sublimation temperature. This high-temperature state occurs as the viscous heating balances radiative cooling around the sublimation temperature of the dust (i.e. the temperature at which 50\% of the dust is sublimated to the gas phase) and is regulated by a thermostat effect, where an increase in temperature would reduce the opacity through sublimation and therefore increase the cooling \citep{ziampras_planet_2026-1}.
The sublimation temperature not only affects the surface density and temperature of the disc during the burst, but also its resulting increase in the accretion rate onto the star and the frequency at which the outbursts occur. From the accretion rate through the inner boundary of the simulation, the accretion luminosity resulting from the events can be calculated. It is given by 
\begin{equation}
    L_{\rm acc} = \frac{G M_\star \dot{M}_{\rm acc}}{R_\star},
\end{equation}
where we assumed a stellar radius of $R_\star = 2.5 R_\odot$. The accretion variability is visualised in Figure \ref{fig:acc_rate}, displaying the accretion rate/luminosity as a function of time for the different setups. The cycle time and increase in accretion rate of the bursts directly correlate with the sublimation temperatures, that is, higher sublimation temperatures lead to stronger but less frequent bursts. Moreover, the setups with low sublimation temperatures clearly do not burst at all and reach a steady accretion rate.

\begin{figure}
    \centering
    \includegraphics[width=\linewidth]{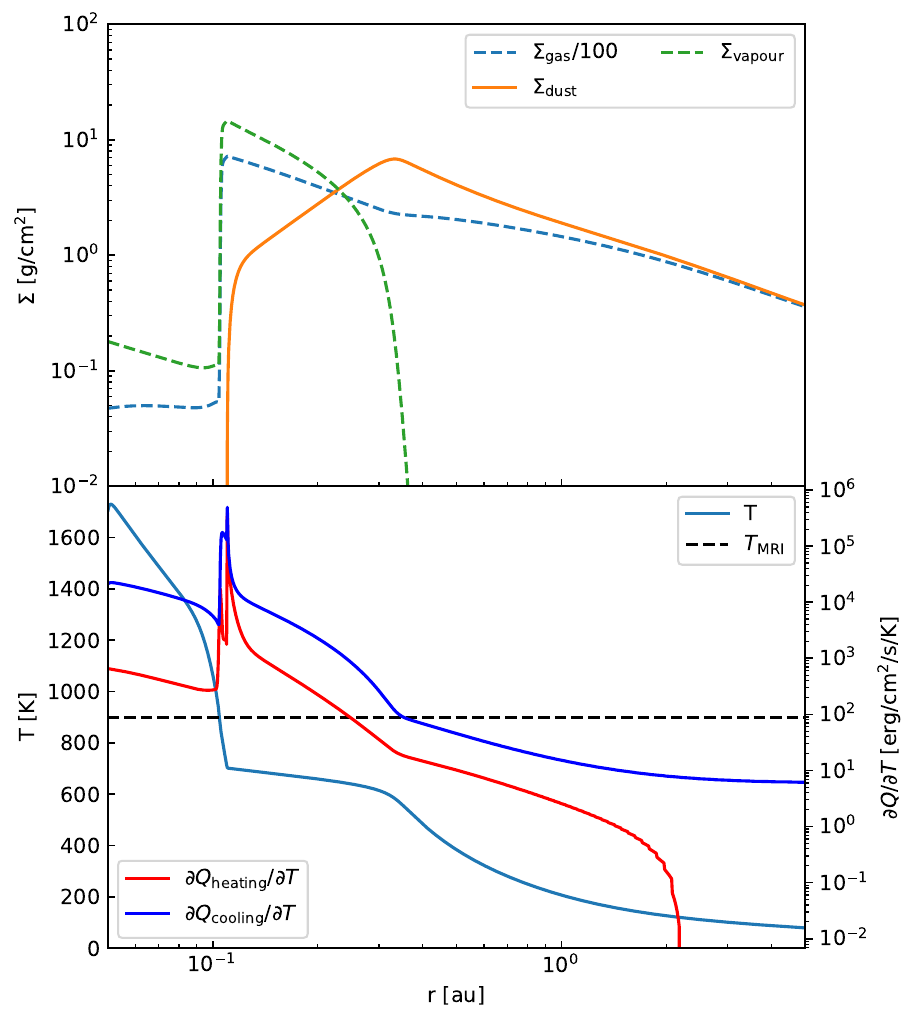}
    \caption{Steady state of the $E_{\rm sub} = 2.1\times 10^4\rm K$ with the gas and dust surface densities (top), the temperature, and the derivative of heating with respect to the temperature (bottom)}
    \label{fig:steady_state}
\end{figure}

The simulations with a binding energy of $E_{\rm sub} \le \SI{26 000}{K}$ evolve to a stable steady state and not periodic outbursts. The structure of the steady state for $E_{\rm sub} = \SI{21 000}{K}$ is shown in Figure \ref{fig:steady_state}. The reason that these simulations reach a stable configuration is clear when we consider the temperature dependence of the heating and cooling terms. To illustrate this, we calculated the derivative of the heating and cooling terms with respect to temperature, as displayed in right axis of the bottom panel of Fig. \ref{fig:steady_state}. The derivatives were calculated numerically in post-processing, that is, $\frac{\partial Q}{\partial t} = \frac{Q(T+\delta) -Q(T)}{\delta}$, but considered the evaporation and condensation to be in equilibrium. They also included the changes in the dust-to-gas ratio (and opacity) as a function of temperature. The instability condition in Eq. \ref{eq:stability_crit} is never met because the cooling derivative becomes significantly higher than the heating derivative around the sublimation temperature of the dust: the aforementioned thermostat effect stabilises against the thermal instability around the MRI activation temperature because the change in cooling as a function of temperature is higher (due to the dust sublimation) than the change in viscous heating around the $\alpha$-transition.

\subsection{Disc parameters}
\label{sec:disk_params}
\begin{figure}
    \centering
    \includegraphics[width=\linewidth]{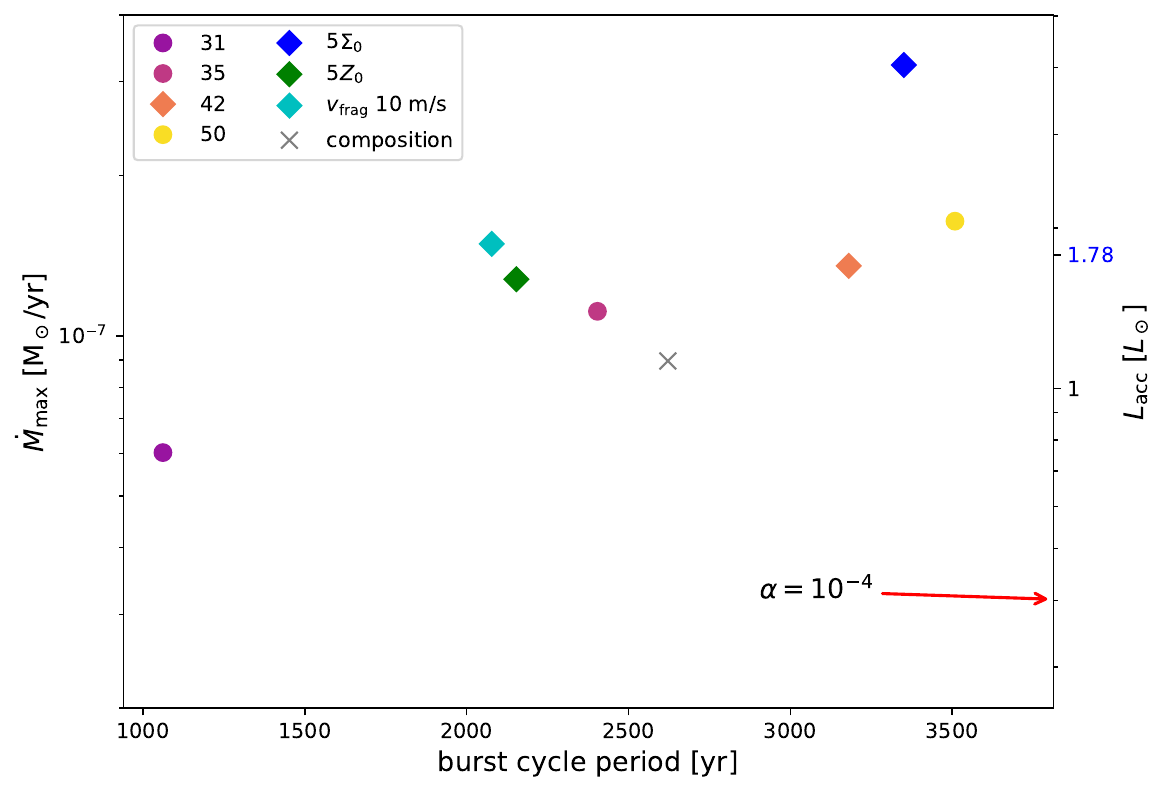}
    \caption{Maximum accretion rate and total burst cycle duration of all the single-component setups (circle: varying sublimation temperature).}
    \label{fig:fisher_plot}
\end{figure}

In addition to the simulations shown above, we wished to explore the effect of other disc parameters on the burst behaviour. The parameters we changed to investigate this were an increased disc mass ($\Sigma_g(1\rm au) = \SI{1000}{g/cm^2}$), a higher disc metallicity ($Z = 0.05$), an increased fragmentation velocity ($v_{\rm frag} = \SI{10}{m/s}$), and a lower dead-zone $\alpha$ ($\alpha_{0} = 10^{-4}$). For each of these parameter changes, we ran a single-component setup with the nominal binding energy of $E_{\rm sub } = \SI{42 000}{K}$. We compare the burst cycle duration and maximum accretion rate of all the aforementioned setups in Fig. \ref{fig:fisher_plot} with the single-component setups above. 

An increase in the disc mass increases the higher maximum accretion rate of the outburst because more mass is flushed onto the star, but it keeps the frequency very similar to the nominal case. This differs from the findings of \cite{cecil_variability_2024}, who reported that the cycle time reduced with increased disc mass, although they only evolved their simulations from an initially forced burst to a second burst, so the timing might still be partially affected by their initial conditions. An increase in metallicity, however, leads to shorter and slightly less potent bursts because with a higher dust-to-gas ratio, the inner edge requires less mass to reach an unstable temperature region through the increased dust-to-gas ratio at the inner dead-zone edge. This follows from equating Eqs. \eqref{eq:Q_cool} and \eqref{eq:Q_visc} and neglecting irradiation,
\begin{equation}
    \Sigma_{\rm gas} \,\tau_{\rm eff} = T^3\,\frac{8\sigma_{\rm sb}\mu}{9 k_{b} \Omega_K \alpha}
    \label{eq:sig_tau},
\end{equation}
and assuming that $\tau_{\rm eff} \propto \epsilon *\kappa_{d}$, the dust-to-gas ratio directly correlates with the temperature.

\begin{figure}
    \centering
    \includegraphics[width=\linewidth]{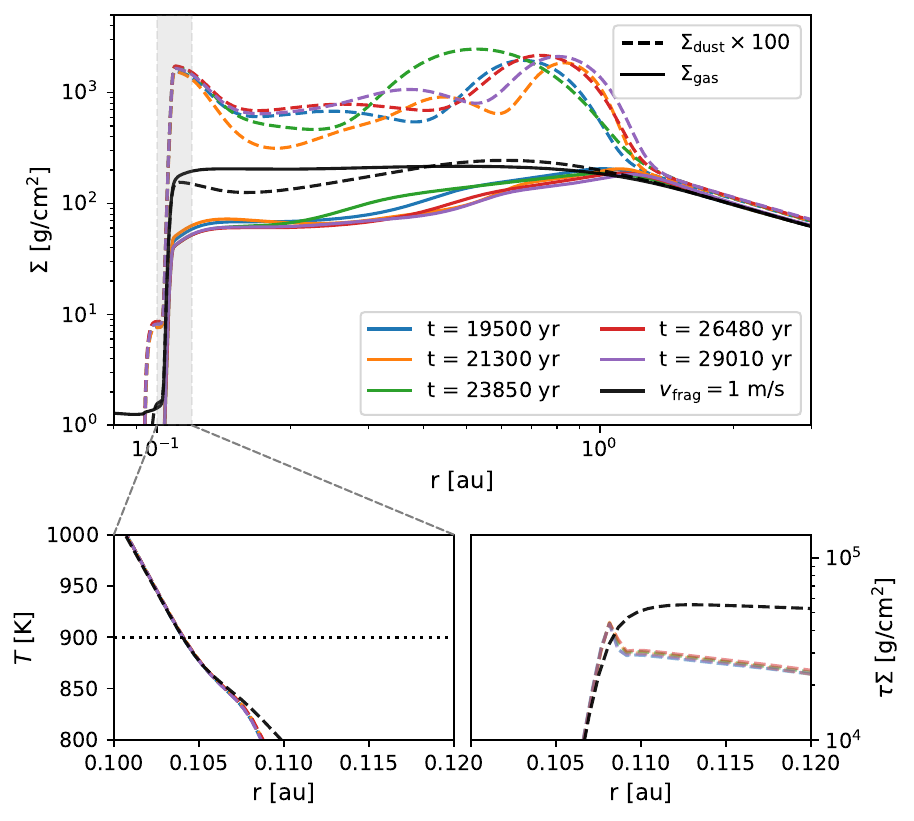}
    \caption{Pre-burst state of the setup with $v_{\rm frag} = 10 \rm m/s$, with the dust and gas surface densities (top), the temperature at the inner edge (bottom left), and the gas surface density times the optical depth (bottom right).}
    \label{fig:10vf_state}
\end{figure}

Unlike the aforementioned cases, the simulations with a high fragmentation velocity did not show a fully periodic outburst behaviour because the gas and dust distributions before subsequent bursts differ from one cycle to the next, as shown in figure \ref{fig:10vf_state}, where we also added the pre-burst state of the nominal simulation as a comparison. Even though the distributions of gas and dust varied from one cycle to the next, the temperature structure at the inner edge where the burst is triggered matches for all of them. The states in which more dust is accumulated at the inner edge clearly require less gas to trigger the burst, as predicted by Eq. \eqref{eq:sig_tau}. To confirm this, we also plot $\tau_{\rm eff}\,\Sigma_{\rm gas}$ for the different pre-burst states. It matches well between the pre-burst states, where the slight differences arise from the contributions from irradiation.

\begin{figure}
    \centering
    \includegraphics[width=\linewidth]{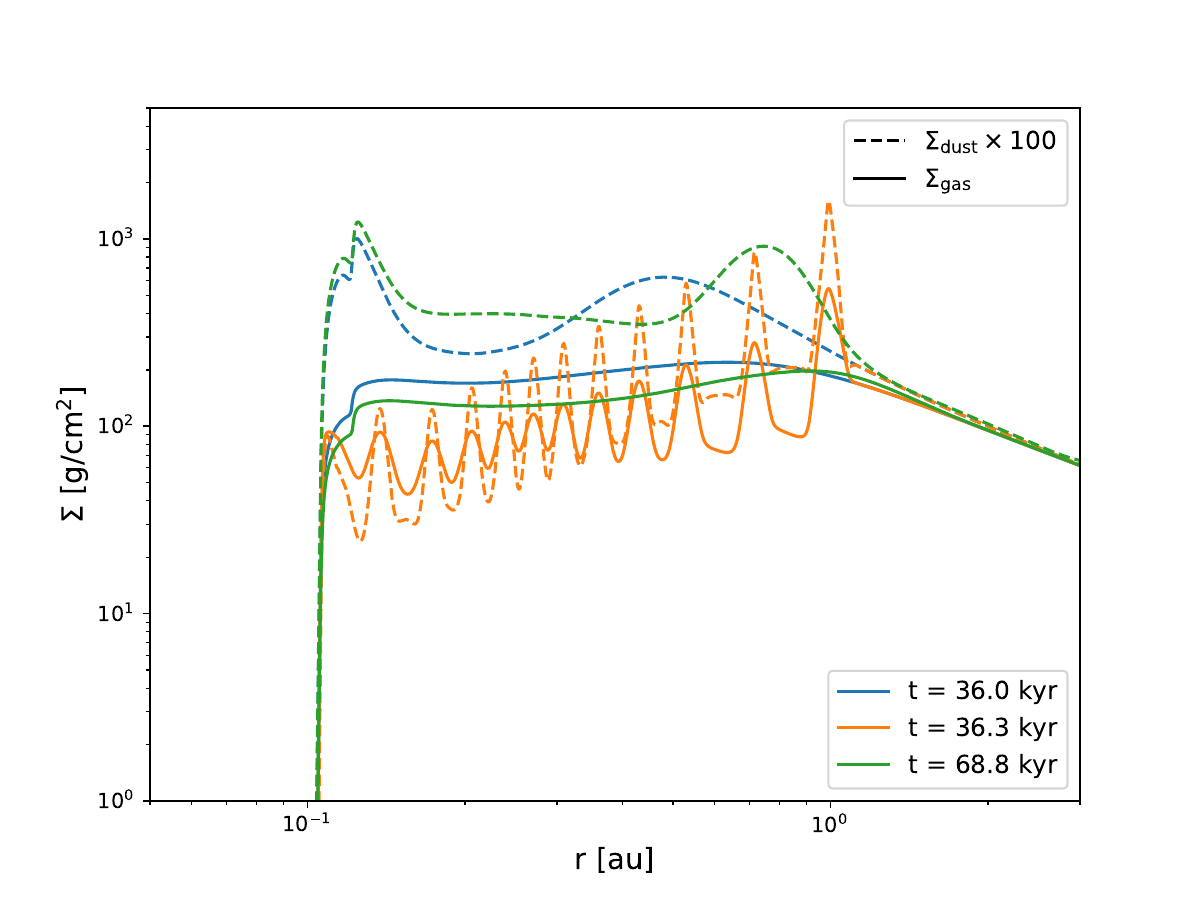}
    \caption{Pre-, post- and pre-burst states of the next cycle for the setup with $\alpha_{\rm DZ} =10^{-4}$, with the gas surface density (solid) and the total dust surface density (dashed). }
    \label{fig:low_alpha}
\end{figure}

We show the pre- and post-burst states of the setup with low turbulence in Fig.~\ref{fig:low_alpha}. The reduced turbulence in the dead zone has the effect that it takes significantly longer to refill the burst region, which was already reported in \cite{ziampras_planet_2026} (also in \cite{bell_using_1994} and \cite{wunsch_two-dimensional_2006} for different setups), leading to a prolonged quiescent phase and a cycle period of $T_{\rm cycle} \sim \SI{36 000}{yr}$. Since the dust has more time to accumulate at the inner pressure trap, which is also less diffusive because the $\alpha$ contrast is higher, this simulation needs to accumulate less gas in the burst region to trigger a burst, leading to a lower spike in accretion rate of $\dot{M}_{\rm max} = \SI{3.2e-8}{M_\odot/yr}$. Additionally, the subsequent burst did not have the exact same pre-burst and post-burst state, but an increased dust-to-gas ratio for the latter one. These results differ from those reported by \cite{ziampras_planet_2026-1}, who estimated significantly longer cycle times ($\sim \SI{100}{kyr}$). The reason might be that they did not consider the dust accumulation in the quiescent phase in their simulations. We did not evolve the simulations for additional bursts, as the steady-state assumption started to break down on timescales where the disc evolution matters, and we started to see a change in pebble flux.

These two simulations nicely show that as the dust starts to decouple from the gas or the time is long enough for it to accumulate, the dust dynamics gain significant importance, leading to outbursts that are no longer strictly periodic (the periods for these setups added in Fig. \ref{fig:fisher_plot} were taken to be the period between the last two bursts). We note that we did not include planetesimal formation in these simulations, which might limit the maximum achievable dust-to-gas ratio \citep{stammler_dsharp_2019}, but did not observe mid-plane dust-to-gas ratios higher than $0.2$. Hence, it remains questionable whether this would affect our results if we considered the planetesimal formation to be triggered at a mid-plane ratio of unity.

\subsection{Full composition}
\label{sec:full_comp}
\begin{figure*}
    \centering
    \includegraphics[width=\linewidth]{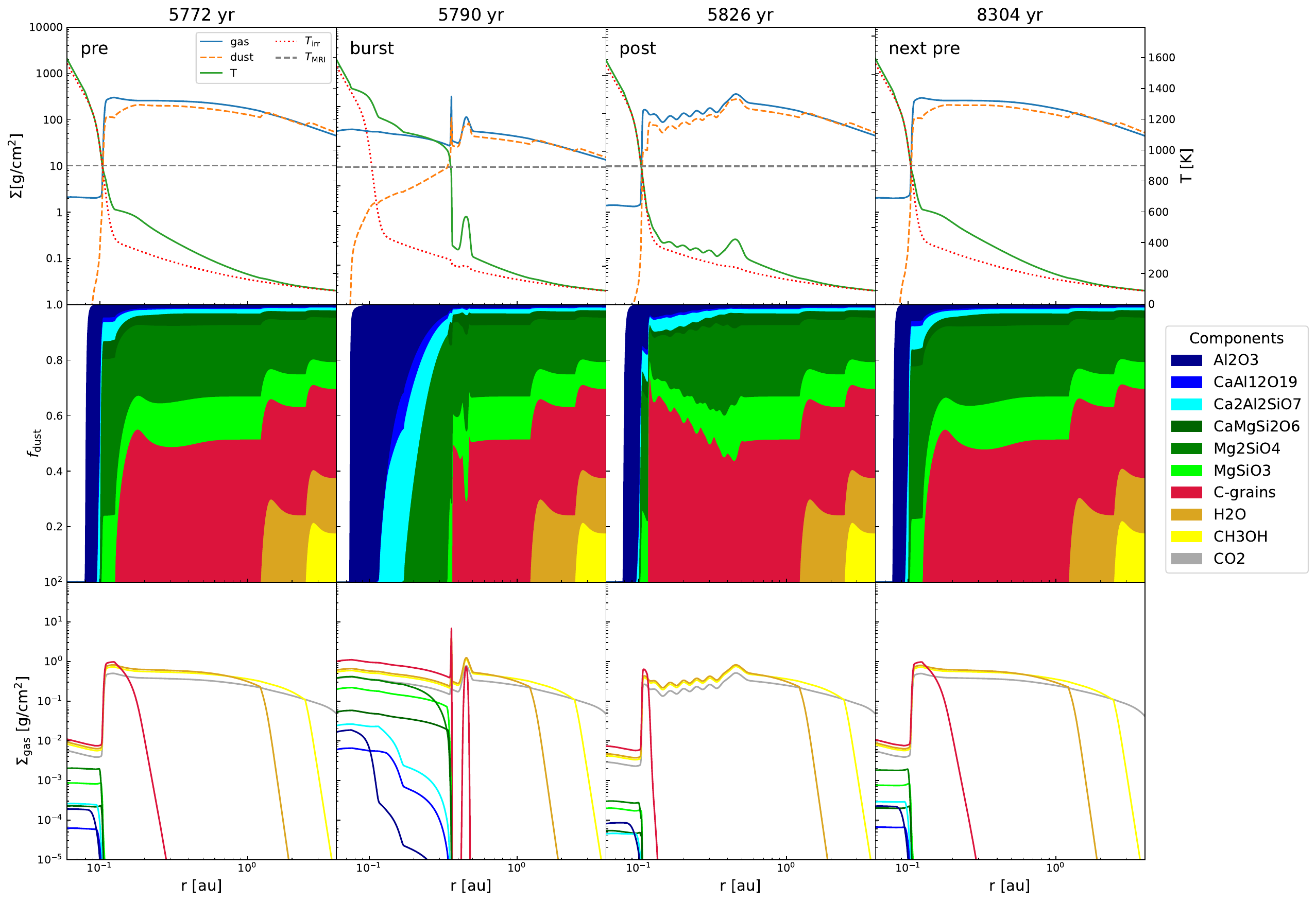}
    \caption{State of the simulation at different times during the simulation (left to right: pre-burst, during, post-, and pre-burst of the next cycle), namely gas (blue) and dust (orange) surface-density and temperature (green) in the top panel, the relative fraction of each component in the solid dust (middle), and the surface distribution of the gas components without H-He (bottom)}
    \label{fig:compo_state}
\end{figure*}

To probe the effect of the outbursts on the composition of dust and gas in the inner disc, we ran a model with the condensation sequence described in Sec. \ref{sec:dust_comp}. We show in Figure \ref{fig:compo_state} the simulation during the different stages of the burst cycle (before, during, and after the burst, and before the next cycle) characterised by its global properties, that is, the total gas and dust surface density and temperature (top), the relative fraction of each component in the solids (middle), and the surface density of all non-H/He components in the gas phase (bottom). 

In the pre-burst state, the condensation sequence of the different elements and the associated increased dust-to-gas ratio are clearly visible when the respective ice lines are passed. Additionally, the abundance of each component increases outside its respective sublimation line (e.g. the carbon grains at $\sim 0.2$ au), as predicted by the cold-finger effect \citep{drazkowska_planetesimal_2017}. 
During the burst, all the innermost components up to carbon evaporate and therefore enrich the gas with silicates and soot up to $\sim 0.5$ au. The temperature structure during the burst state shows multiple steps that are caused by the thermostat effect of the different respective components that dominate the opacities at different radii. For instance, the transition at which the Al-bearing species control the opacities (up to 0.2 au) and then the temperature drops as the Mg-bearing silicates are present. The burst does not travel out far enough to affect any of the volatiles. In the post-burst state, most of the material recondensed onto the solids faster than it was accreted onto the star in the gas phase, even though the vast majority of material was evaporated in the burst. However, the radial distribution of the solids was significantly altered (e.g. C-grains). Nevertheless, the subsequent viscous evolution of the quiescent phase returned the composition to the identical pre-burst state as before. 

While this suggests that the outburst only temporarily affects the composition of the inner disc, it prevents a significant pile-up at the sublimation lines of the silicates and carbon grains, which would be expected at a stable inner rim.

\section{Discussion}
\label{sec:discussion}
\subsection{Model limitations}
Our 1D approach using a parametric dust evolution model comes with a few caveats and shortcomings that have to be discussed. Firstly, treating the outbursts with our 1D viscous solver neglects effects that could be captured by a full Riemann solver. The different treatment of the hydrodynamics leads to slight differences in the dust distribution of the dust and gas at the inner disc edge, where our code proves to be slightly more diffusive than previous works (see Figure 1.D. in \cite{ziampras_planet_2026-1}). Additionally, the formation of vortices in the burst front observed in 2D simulations \citep{ziampras_planet_2026,cecil_mri-triggered_2026} cannot be captured with our approach. Nevertheless, the general outburst structure remains the same in 1D and 2D, with the most significant change being the sharpness of the ring features in the post-burst state at low viscosities, which is thought to not affect our results as we do not observe significant dust trapping around these structures in any of our models in the first place.

A few caveats with respect to the treatment of the inner rim have to be mentioned. Firstly, the treatment of the irradiation angle $\theta$ as a purely scale-height-dependent variable might not hold, as the directly irradiated inner ring can puff up and cast a shadow on the outer regions of the disc \citep{dullemond_inner_2010,flock_effect_2025}. This would reduce the heating term $Q_{irr}$ in Eq. \eqref{eq:Q_irr} for the shadowed regions, but since the burst dynamics are dominated by viscous heating and radiative cooling, the effects on the burst behaviour are expected to be secondary. Additionally, it is not fully self-consistent to model the directly irradiated inner rim by modifying the irradiation angle at a static radius because the inner rim is in principle given by the optical surface with respect to the direct irradiation of the star, which is also a function of scale height \citep{ueda_analytic_2017}. For a more consistent treatment, the optical surface would need to be calculated dynamically by ray-tracing the direct stellar irradiation and adjusting $r_{\rm rim}$ accordingly; however, this would require a more detailed treatment of the gas opacities \citep{cecil_role_2026}, which is beyond the scope of this work.

To limit the complexity while still capturing the general physical behaviour, we omitted iron in the models, because its sublimation coincides with that of enstatite and forsterite and it is less abundant than these minerals. The addition of iron to the model would shift the burst period and accretion rate to slightly shorter and higher values, towards the model with a parametrised sublimation line at 35 K (which closely represents the enstatite, fosterite, and iron sublimation energy).

In the treatment of the composition, we ignored the effects of chemistry, which has been shown to affect the composition of the inner disc \citep{booth_planet-forming_2019,sellek_chemical_2025,molyarova_metamorphoses_2026} even though the compositional budget is still largely determined by dust dynamics. The predictions of many of these models showed variations in the inner disc chemistry, which is usually represented by the C/O ratio, which varies on timescales of $10^5-10^6 \rm yr$. This is much longer than our run times. Additionally, we assumed that sublimation and evaporation are reversible processes, which is not necessarily true \citep{houge_burned_2025}, as the components, especially the carbon grains, would readily interact in the gas phase into more volatile components. This is expected to play an important role in the carbon budget of the inner disc because it would prevent the recondensation of carbon-rich volatiles onto the dust. This might also affect future burst cycles.

In our model, the sublimation energy of different species is not a function of the changing pressures, but represents a single pressure of $10^{-4}$ atm \citep[for details see][]{yoneda_condensation_1995}, which would approximately be the equivalent to the pressure of a disc at 0.5 au with a surface density of 2000~g$\,$cm$^{-2}$. In a more self-consistent model, the sublimation energy needs to dynamically adjust to the current density and temperature integrated in the model. The increasing temperature and density leads to increasing pressures, and thus, sublimation temperatures by up to $\sim5\%$.

We focused on the burst process and the effect of sublimation and condensation of the minerals on the burst physics. We included more volatile species and refractory carbon carriers to understand whether the burst affects the composition at these parts of the disc, however. For the refractory carbon, we approximated the general group of polycyclic aromatic hydrocarbons and other refractory long-chain hydrocarbons by matching the mean molecular weight of C$_4$H$_{10}$ in the gas phase, because refractory carbons do not just sublimate, but are often chemically processed into shorter organics such as C$_4$H$_{10}$ \citep{gail_radial_2001}. Our current model only approximates the effect as a single carbon refractory species without chemical processing. Our results show that especially the refractory hydrocarbons are affected by the burst even if their sublimation temperature is too low to cause the burst. These chemical processes will be followed up in future works.

\subsection{Consequences for disc composition}
The subject of several recent JWST/MIRI observations and programs has been to probe the inner disc chemistry \citep[e.g.][]{banzatti_jwst_2023,grant_transition_2025,henning_minds_2024,mcclure_refractory_2025}. The combination of line observations with thermochemical models allows us to infer elemental abundance ratios such as C/O \citep{woitke_modelling_2018,anderson_observing_2021}. The inner disc C/O ratio has been proposed as a tool for constraining the dust evolution in protoplanetary discs as it is strongly affected by the pebble flux to the inner disc, and therefore, the presence of substructures \citep{mah_close-ice_2023,sellek_chemical_2025}, and because serves as a good indicator for understanding the chemistry/composition of the inner disc. Theoretical models explaining the inner disc chemistry have largely focused on radial transport and chemical evolution to explain the inner disc composition as a result of pebble drift and chemical processing (e.g.  \citep{booth_chemical_2017,sellek_co2_2025,houge_burned_2025,molyarova_metamorphoses_2026,mah_close-ice_2023}). The outbursts presented here indicate an additional complication to these predictions, as the carbon released in the outbursts, combined with the potential non-reversibility of the condensation of the carbon grains \citep{gail_radial_2001,li_earths_2021,houge_burned_2025}, might increase the C/O ratio in the inner disc at earlier times. However, to gain a clear picture of the additional compositional imprint of the outbursts, our simulations should be extended to also include the outer disc and longer run times. This would determine how long the outbursts persist and how they interact with the changing composition of solids arriving in the inner disc.

\subsection{Comparison with variability in young stellar objects}
Our simulations show accretion luminosity variations of order unity on kilo-year timescales, which do not resemble the typical outburst behaviour of young stellar objects \cite{fischer_accretion_2023}. However, we explored several parameters that strongly shift the maximum outburst accretion rate and cycle frequency. For example, the maximum accretion rate increased for an enhanced disc mass, and the cycle duration is strongly dependent on $\alpha_{\rm DZ}$. While the model parameters might be tuned to fit EX-Lupi-type events \citep{herbig_ex_2007,fischer_accretion_2023} that show an increase of several magnitudes lasting up to a few years, it is not possible to match FU-Ori-type bursts with our model (they display a luminosity increase of $\sim 100 L_\odot$). This is consistent with previous works, as theoretical FU-Ori models, for example \citep{zhu_nonsteady_2009,bae_variable_2013,bae_accretion_2014}, relied on much higher surface densities that are consistent with younger-type discs (and gravitationally driven $\alpha_{\rm DZ}$) to achieve accretion luminosities of $\sim 10^2 -10^3 L_\odot$, which agrees better with observations of FU-Ori-type events. These simulations must also take several physical processes into account that are not accounted for in our models and become relevant for these massive early-type disks, such as gravitational (in-)stability \citep{zhu_nonsteady_2009} and gas opacities suited for higher temperatures \citep{bell_using_1994}. 

\section{Conclusion}
\label{sec:conclusion}
We presented 1D models of stellar outbursts driven by a thermal instability at the inner edge of the dead zone. For the first time, we included the concurrent coagulation, evaporation, and condensation of dust and the volatile species enveloping dust grains. This allowed us to track the composition of the inner disc during these outbursts using our publicly available \texttt{TriPoDpy} code. We used these models to explore the effect of evaporation and condensation on the dynamics of these outbursts, and in turn, the effect of the outbursts on the composition of the inner disc. From our models with a single dust sublimation temperature, we found that
\begin{itemize}
    \item the frequency and maximum accretion rate of the outbursts strongly depend on the dust sublimation temperature, with higher sublimation temperatures leading to less frequent outbursts with higher maximum accretion rates/luminosities, and
    \item for a very low dust binding energy ($E_{\rm sub} \le \SI{2.6e4}{K}$), the disc becomes stable and does not burst because of the thermostat effect.
    \item We explored the effect of different disc parameters on the dynamics of the bursts, identifying the dust accumulation and the inner edge (in addition to the mass) as a prime determining factor to trigger bursts. 
    \item For certain setups, the bursts are not strictly periodic as the dust accumulation from drift changes the burst region over time.
\end{itemize}
When modelling the dust with a full condensation sequence to track the changes in composition caused by the outbursts, we found that 
\begin{itemize}
    \item the outburst sublimates the dust out to 0.5 au and enriches the gas temporarily with silicates and soot, and
    \item the compositional imprint of the outburst is reset by the viscous evolution in the quiescent phase between outbursts.
\end{itemize}

Several open questions remain regarding the connection between the inner disc composition and outbursts. As the inner disc chemistry has been shown to be highly dependent on stellar mass \citep{grant_transition_2025} showing a tentative trend in C/O ratio, it would be worth exploring whether the outbursts have a different effect on the composition of lower-mass stars. Additionally, it would be important to consider the non-reversibility of carbon recondensation, as suggested in \cite{houge_burned_2025}, which will be followed up on in future works.

\begin{acknowledgements}
NK, TB, AP and AZ knowledge funding from the European Union under the European Union’s Horizon Europe Research and Innovation Programme 101124282 (EARLYBIRD). Views and opinions expressed are those of the authors only. TB acknowledges funding by the Deutsche Forschungsgemeinschaft (DFG, German Research Foundation) under Germany’s Excellence Strategy - EXC-2094 - 390783311. RAB thanks the Royal Society for their support via a University Research Fellowship. AR is supported by the Science and Technology Facilities Council (STFC) award ``Building Rocky Planets in the Inner Disc: From Dust to Planetesimals" (project reference UKRI1189).
\end{acknowledgements}

\section{Data availability}
Data from our simulations is shared upon reasonable request.

\bibliographystyle{aa}
\bibliography{references}

\appendix
\nolinenumbers
\FloatBarrier
\section{Comparison with \cudisc}
\label{ap:cudisc}
To show how well our composition module performs, we run a test setup of the CO ice-line as described in \cite{stammler_redistribution_2017} and compare our results with the same setup performed with \cudisc \cite{robinson_introducing_2024}, which includes a treatment of evaporation and condensation onto a fully resolved distribution of particle sizes. The methods employed for ice-vapour chemistry in \cudisc are described in \cite{robinson_co_2026}; the 1D version used for the comparison detailed here includes the same vertical integration factor in the adsorption rate used by \tripodpy.
The shared parameters and initial conditions for these simulations can be seen in Table \ref{tab:CO_setup}. To isolate the effect of evaporation and condensation and the associated dynamics we kept the mean molecular weight and the bulk density of the dust constant regardless of composition. We initialise the dust with an MRN-like size distribution and all the CO in the gas phase.

\begin{table}[h]
    \centering
    \caption{Parameters for the CO-ice-line comparison}
    \begin{tabular}{c c c}
     Name &  Parameter  & Value \\
       \hline
       \hline
      Initial maximal grain size & $a_{\rm max, ini}$  & $10^{-4}$ cm \\
      Minimal grain size & $a_{\rm min}$  & $10^{-5}$ cm \\
      Initial size power law & $q_{\rm ini}$ & -3.5 \\
    Dust bulk density & $\rho_{\rm dust} = \rho_{\rm CO}$  & $1.6 \rm  g/cm^3$ \\
      Temperature &  T & $150 \left( \frac{R}{\rm au}\right)^{-0.5}$ K \\
      Initial gas surface density & $\Sigma_{\rm g,ini}$ & $1000\left( \frac{R}{\rm au}\right)^{-1} \rm g/cm^2$ \\
      Mean molecular weight & $\mu$ & $2.4 m_{p}$\\
      Initial dust to gas ratio & $\varepsilon$ & $10^{-2}$ \\
      \hline
      Initial CO surface density & $\Sigma_{\rm CO}$ & $\Sigma_{\rm g,ini}/700$ \\
      Sublimation energy CO & $E_{\rm sub, CO}$ & $850 \rm K$ \\
      Trial frequency CO & $\nu_{\rm CO}$ & $7 \times10^{11} \rm Hz$ \\
    \end{tabular}

    \label{tab:CO_setup}
\end{table}

\begin{figure}
    \centering
    \includegraphics[width=\linewidth]{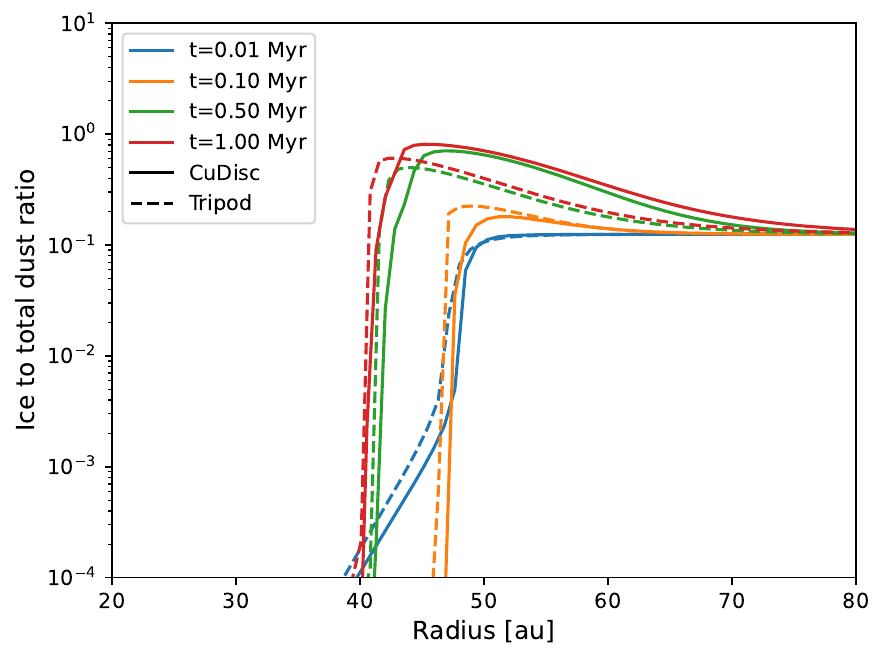}
    \includegraphics[width=\linewidth]{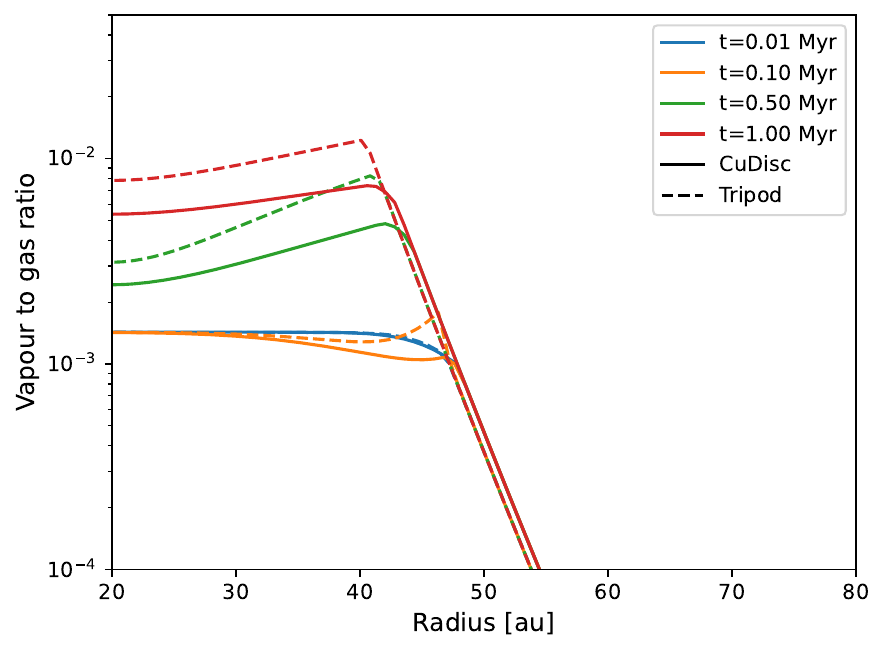}
    \caption{The CO-ice to dust (top) and CO-vapour to gas ratio in the \tripodpy and \cudisc throughout the simulation}
    \label{fig:ice_ratios}
\end{figure}

\begin{figure*}
    \centering
    \includegraphics[width = \linewidth]{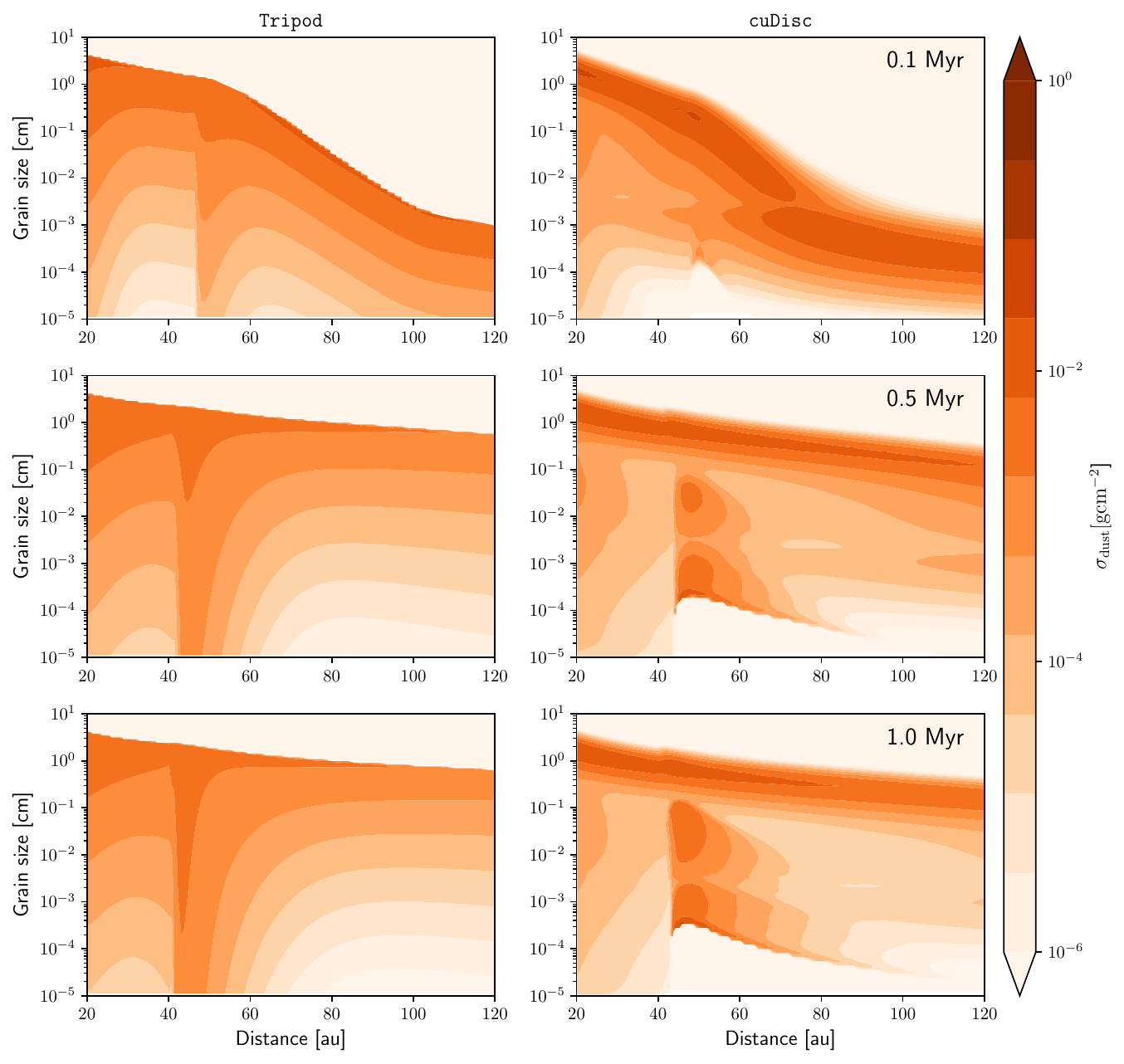}
    \caption{The dust size distribution (including CO-ice), with \tripodpy (left) and \cudisc (right) at different times}
    \label{fig:size_dist_cudsic}
\end{figure*}

When we look at the size integrated CO to dust and gas ratios, shown in Fig. \ref{fig:ice_ratios}, the CO concentration across the ice line is consistent between the two simulations. The sharper concentration peaks in \tripod are caused by less of the total ice fraction residing on the smallest grains. In Figure \ref{fig:size_dist_cudsic}, we show the total dust size distribution (dust and CO-ice) of the two simulations at different snapshots. The codes have two main differences. Firstly, in \cudisc, the minimum size of dust grains is shifted when ice is deposited on the dust grains. We neglected this effect in our code. Secondly, around the ice line, the size distribution becomes multimodal in \cudisc, which can not be fully represented by a power law. 

\FloatBarrier
\section{\tripod method}
In this section, we describe the \tripod method \citep{pfeil_tripod_2024} in more detail.
Starting from the assumed particle size distribution of $n\left(a\right)\propto a^{q}$ with a fixed minimal size $a_{\rm min}$ and maximum size $a_{\rm max}$, we derive the parametric description given by $\Sigma_0$, $\Sigma_1$ and \amax. The mass-density distribution follows $\sigma\left( a\right) \propto n\left(a\right) m_p(a) \propto a^{q+3}$. From integration and normalisation to the total dust surface density, we obtain

\begin{align}
\Sigma_{tot} &= \int\limits_{a_{\rm min}}^{a_{\rm max}} \sigma\left(a\right) \mathrm{d}a = \int\limits_{a_{\rm min}}^{a_{\rm max}} Ca^{q+3} \mathrm{d}a = \left\{ \begin{aligned} \frac{C}{q+4} \left(a_{\rm max}^{q+4} - a_{\rm min}^{q+4}\right) \textrm{for }q\neq -4\\ C \left[\log(a_{\rm max}/a_{\rm min})\right] \textrm{for }q = -4 \end{aligned} \right. \\
&\Rightarrow \sigma\left(a\right) = \left\{ \begin{aligned} \frac{\Sigma_{tot}(q+4)}{a_{\rm max}^{q+4} - a_{\rm min}^{q+4}} a^{q+3}\textrm{for }q\neq -4 \\ \frac{\Sigma_{tot}}{\log(a_{\rm max}/a_{\rm min})}a^{q+3} \textrm{for }q = -4 \end{aligned} \right.
\end{align}
Therefore, the total surface density in the mass interval $[a_0, a_1]$ is,
\begin{align}
\Sigma_{[a_0, a_1]} = \int\limits_{a_0}^{a_1} \sigma\left(a\right) \mathrm{d}a = \left\{ \begin{aligned} \Sigma_{tot} \frac{a_1^{q+4}-a_0^{q+4}}{a_{\rm max}^{q+4}-a_{\rm min}^{q+4}}\textrm{for }q\neq -4 \\ \Sigma_{tot} \frac{\log(a_1/a_0)}{\log(a_{\rm max}/a_{\rm min})} \textrm{for }q = -4 \end{aligned} \right.
\end{align}
This allows us to define the small and large mass bin $\Sigma_0 = \Sigma_{[a_{\rm min},a_{\rm int}]}$ and $\Sigma_1 = \Sigma_{[a_{\rm int},a_{\rm max}]}$ with $a_{\rm int} = \sqrt{a_{\rm max} a_{\min}}$. We can now rewrite the power law exponent $q$ as a function of our three chosen parameters as,

\begin{align}
\log\left(\frac{\Sigma_1}{\Sigma_0}\right) &= \log\left(\frac{a_{\rm max}^{q+4}-a_{\rm int}^{q+4}}{a_{\rm int}^{q+4}-a_{\rm min}^{q+4}}\right) \\ 
&= \left(q+4\right) \log\left(\frac{a_{\rm max}}{a_{\rm int}}\right) + \log \left[\frac{1-\sqrt{\frac{a_{\rm min}}{a_{\rm max}}}^{q+4}}{1-\sqrt{\frac{a_{\rm min}}{a_{\rm max}}}^{q+4}}\right]\\
&\Rightarrow q = \frac{\log\left(\Sigma_1/\Sigma_0\right)}{\log\left(a_{\rm max}/a_{\rm int}\right)} - 4
\end{align}
where the case $q =-4$ follows trivially.
\subsection{Evolution}
The evolution of the dust distribution through mutual collisions and the interactions with the gas disc is described in the following way.
The growth of the maximal particle size  \amax is described by the mono-disperse growth rate as follows,
\begin{align}
\dot{a}_{max} = \frac{\Sigma_{1} \Delta v_{\max}}{\rho_{m} \sqrt{2 \pi} H_{1}}\left(\frac{1 - \left( \frac{v_{\mathrm{frag}}}{\Delta v_{\max}} \right)^{s}}{1 + \left( \frac{v_{\mathrm{frag}}}{\Delta v_{\max}}\right)^{s}}\right),
\end{align}

where $\Delta v_{max} = \Delta v{(0.4 a_{\rm max}, a_{\rm max})}$, is the relative velocity between particles of size $a_{\rm max}$ and particles that are 0.4 times their size. $s>0$ is a number derived from calibration ($s=3$). The relative velocities between the particles take into account contributions from the turbulence, settling, relative drift motion and Brownian motion as described in \cite{stammler_dustpy_2022}.

To calculate the radial gas transport, we use the mass-averaged particle size in each bin to evaluate the radial velocities:

\begin{align}
\langle a\rangle_{[b,c]} = \frac{\int\limits_{b}^{c}a \sigma\left(a\right) \mathrm{d}a}{\int\limits_{b}^{c} \sigma\left(a\right) \mathrm{d}a} 
\end{align}

we solve the diffusion-advection equations with the radial velocities ($v_i = v(a_i)$) and diffusivities ($D_i$) as in \cite{stammler_dustpy_2022} for the sizes $a_0 = \langle a\rangle_{[a_{\rm min},a_{\rm int}]}$ and $a_1 = \langle a\rangle_{[a_{\rm int},a_{\rm max}]}$ respectively:
\begin{equation}
\frac{\partial}{\partial t}\Sigma_{i}
+ \frac{1}{r}\frac{\partial}{\partial r}
\left( r \Sigma_{i} v_{i} \right)
- \frac{1}{r}\frac{\partial}{\partial r}
\left[
r D_i \Sigma_g \frac{\partial}{\partial r}
\left( \frac{\Sigma_{i}}{\Sigma_g} \right)
\right]= \left .\frac{\partial  \Sigma_i}{\partial t}\right\vert_{\rm coag}
\end{equation}

To solve for the evolution of \amax we treat it as a passive scalar of $\Sigma_1$ with the additional growth term $\dot{a}_{\rm max}$
\begin{align}
\frac{\partial}{\partial t}(\Sigma_{1}{a}_{\rm max})
+ \frac{1}{r}\frac{\partial}{\partial r}
\left( r \Sigma_{1}{a}_{\rm max} v_{1} \right)
- \frac{1}{r}\frac{\partial}{\partial r}
\left[
r D_1 \Sigma_g \frac{\partial}{\partial r}
\left( \frac{\Sigma_{1}{a}_{\rm max}}{\Sigma_g} \right)
\right] \\= \dot{a}_{\rm max} \Sigma_1 + a_{\rm max}\left .\frac{\partial  \Sigma_1}{\partial t}\right\vert_{\rm coag}\nonumber
\end{align}
The source term $\left .\frac{\partial  \Sigma_i}{\partial t}\right\vert_{\rm coag}$ describes the mass exchange between $\Sigma_0$ and $\Sigma_1$ due to sweep up and fragmentation and ensures that particle size distribution relaxes to the appropriate one. We define two main mechanisms that shift mass between the bins: the sweep up of small particles by the large ones and fragmenting collisions, shifting mass to the small mass bin. These two processes can be described via,

\begin{align}
\dot{\Sigma}_{0 \rightarrow 1} &= \frac{\Sigma_0 \Sigma_1 \sigma_{01} \Delta v_{01}}{m_1 2 \pi H_0 H_1} \int\limits_{-\infty}^\infty \exp\left[-\frac{z^2}{2}\frac{H_0^2 + H_1^2}{H_0^2 H_1^2}\right] \mathrm{d}z \nonumber \\ &= \frac{\Sigma_0 \Sigma_1 \sigma_{01} \Delta v_{01}}{m_1 \sqrt{2 \pi \left(H_0^2 + H_1^2\right)}} \\ \dot{\Sigma}_{1 \rightarrow 0} &= \frac{\Sigma_1^2 \sigma_{11} \Delta v_{11}}{m_1 2 \pi H_1^2} \mathcal{F} \int \limits_{-\infty}^\infty \exp \left[-\frac{z^2}{H_1^2}\right]\mathrm{d}z \nonumber \\ &= \frac{\Sigma_1^2 \sigma_{11} \Delta v_{11}}{m_1 2 \sqrt{\pi} H_1} \mathcal{F}
\end{align}

where $\sigma_{mn} = \pi(\langle a_m\rangle + \langle a_m\rangle)^2$ and $\Delta v_{mn}$ are the collision cross-section and typical relative velocity between mass bins $m$ and $n$, where $\Delta v_{10} = \Delta v (\langle a_1\rangle,\langle a_0\rangle)$ and $\Delta v_{10} = \Delta v (\langle a_1\rangle,0.4\langle a_1\rangle)$ whre the factor $0.4$ is a calibration factor. The term $\mathcal{F}$ is found in the equilibrium by equating $\dot{\Sigma}_{d, 0 \rightarrow 1}$ and $\dot{\Sigma}_{d, 1 \rightarrow 0}$:

\begin{align}
\frac{\Sigma_1}{\Sigma_0} = \sqrt{\frac{2H_1^2}{H_0^2 + H_1^2}} \frac{\sigma_{01}\Delta v_{01}}{\sigma_{11}\Delta v_{11}} \mathcal{F}^{-1}.
\end{align}
Using the definition of $q$ we find,
\begin{align}
\label{eq:F_mathcal}
\mathcal{F} = \sqrt{\frac{2H_1^2}{H_0^2 + H_1^2}} \frac{\sigma_{01}\Delta v_{01}}{\sigma_{11}\Delta v_{11}}\left(\frac{a_{\rm max}}{a_{\rm min}}\right)^{-\left(q'+4\right)},
\end{align}

where $q'$ is the equilibrium power law exponent, the model relaxes to. $q'$ depends on the velocity regime, where $q'_{frag} \in [-3.5,-3.75]$ in the fragmentation limit and $q`_{stick} = -3.0$ out of equilibrium. We assign a transition function $f$ to smoothly change from growth to fragmentation when the relative velocities reach the fragmentation velocity $v_{frag}$. It has to satisfy,

\begin{align}
f\left(\Delta v_{11} / v_{frag}\right) = \left\{ \begin{aligned} \rightarrow 1 \textrm{ for } \Delta v_{11} \geq v_{frag} \\ \rightarrow 0 \textrm{ for } \Delta v_{11} \ll v_{frag} \end{aligned} \right.
\end{align}
and therefore $q' = f*q'_{\rm frag} + (1-f)q'_{\rm stick}$. The transition function and the exact definition of $q'_{frag}$ can be found in \cite{pfeil_tripod_2024}.

The timestep is calculated in the following way: we calculated the explicit source terms of $\Sigma_{i}$ due to coagulation and radial transport and limit the timestep so they are not allowed to change by more than $10\%$. The changes caused by evaporation and condensation were not considered here because they occur on restrictively short timescales and lead to an equilibrium state.
Additionally, we calculated the coagulation source term and changes in $a_{\rm max}$ due to advection with the gas and limit the relative change in $a_{\rm max}$ to $10\%$.

\label{ap:tripod}
\end{document}